\documentclass[12pt]{article}
\renewcommand{\arraystretch}{1.5}
\renewcommand{\bar}{\overline}
\newcommand{\tr}{\mathop{\rm tr}}

\newcommand{\half}{{1\over 2}}

\input prepictex
\input pictex
\input postpictex
\newdimen\tdim
\tdim=\unitlength
\def\stpltsmbl{\setplotsymbol ({\small .})}

\def\photonrd #1 #2 *#3 /{\multiput {\copy\phrd}  at
#1 #2 *#3 10 0 /}
\newbox\phdr
\setbox\phdr=\hbox{\beginpicture
\setcoordinatesystem units <\tdim,\tdim>
\stpltsmbl
\setquadratic
\plot
0 0
3 -2.5
0 -5 
-3 -7.5 
0 -10 
/
\endpicture}
\def\photondr #1 #2 *#3 /{\multiput {\copy\phdr}  at
#1 #2 *#3 0 -10 /}

\newcommand{\beq}{\begin{equation}}
\newcommand{\eeq}{\end{equation}}

\newcommand{\laem}{\stackrel{<}{\sim}}
\newcommand{\gaem}{\stackrel{>}{\sim}}

\begin{document}
\begin{titlepage}
\def\thepage {}        

\title{Effective Field Theory of Vacuum Tilting}

\author{
\renewcommand{\arraystretch}{1}
\begin{tabular}{c}R. Sekhar Chivukula\thanks{sekhar@bu.edu}\\
Physics Department \\
Boston University \\
Boston MA  02215\end{tabular}
\and
\renewcommand{\arraystretch}{1}
\begin{tabular}{c}Howard Georgi\thanks{georgi@physics.harvard.edu}\\
Lyman Laboratory of Physics\\
Harvard University\\
Cambridge, MA 02138\end{tabular}}

\date{8/10/98}
\maketitle

\bigskip
\vspace{-12pt}
\begin{picture}(0,0)(0,0)
\put(355,250){BUHEP-98-15}
\put(355,235){HUTP-98/A042}
\end{picture}
\vspace{24pt}

\begin{abstract}
  Simple models of topcolor and topcolor-assisted technicolor rely on a
  relatively strong $U(1)$ gauge interaction to ``tilt'' the vacuum.
  This tilting is necessary to produce a top-condensate, thereby
  naturally obtaining a heavy top-quark, and to avoid producing a
  bottom-condensate. We identify some peculiarities of the
  Nambu-Jona-Lasinio (NJL) approximation often used to analyze the
  topcolor dynamics. We resolve these puzzles by constructing the
  low-energy effective field theory appropriate to a mass-independent
  renormalization scheme. We construct the power-counting rules for such
  an effective theory. By requiring that the Landau pole associated with
  the $U(1)$ gauge theory be sufficiently above the topcolor gauge boson
  scale, we derive an upper bound on the strength of the $U(1)$
  gauge-coupling evaluated at the topcolor scale. The upper bound on the
  $U(1)$ coupling implies that these interactions can shift the
  composite Higgs boson mass-squared by only a few per cent and,
  therefore, that the topcolor coupling must be adjusted to equal the
  critical value for chiral symmetry breaking to within a few per cent.

\pagestyle{empty}
\end{abstract}
\end{titlepage}

\setcounter{section}{0}

\section{Introduction}
\setcounter{equation}{0}

Simple models of topcolor \cite{Hill:1991at} and topcolor-assisted
technicolor \cite{Hill:1995hp} rely on chiral-symmetry breaking driven
by the combination of an isospin-symmetric top-color gauge interaction
and a relatively strong isospin-violating $U(1)$ gauge interaction which
couple to the third generation of quarks. The top-color gauge
interaction binds a composite Higgs boson which, if the interaction is
near critical, is very light compared to its compositeness scale and
thus can be described in an effective Lagrangian description with a
fundamental scalar
field~\cite{Kaplan:1984fs,Kaplan:1984sm,Dugan:1985hq,Miranskii:1989ds,Miranskii:1989xi,Nambu:1989jt,Marciano:1989xd,Bardeen:1990ds}.
The $U(1)$ interaction is necessary to ``tilt'' the vacuum and produce a
top-condensate, thereby naturally obtaining a heavy top-quark, but to
avoid producing a bottom-condensate.  We begin in section 2 by
identifying some peculiarities of the Nambu-Jona-Lasinio (NJL)
\cite{Nambu:1961er} approximation often used to analyze the
chiral-symmetry breaking dynamics. In section 3, we construct the
low-energy effective field theory appropriate to a mass-independent
renormalization scheme and construct the power-counting rules for such
an effective theory.  The low-energy effective theory allows for an
analysis of the dynamics of chiral-symmetry breaking beyond the NJL
approximation, and the puzzles found in that approximation are resolved.

Phenomenological constraints \cite{Hill:1994} require that the
mass of the topcolor gauge bosons,
$M$, be substantially larger than the top-quark mass. For this to
be possible, the topcolor chiral phase transition must be (at least
approximately) second order \cite{Chivukula:1990bc} and the topcolor
coupling strength (renormalized at scale $M$) must be adjusted to be
close to the critical value $\alpha_c$ for chiral symmetry breaking. To
tilt the vacuum, the $U(1)$ coupling must be strong enough to prevent
bottom-quark condensation.  In section 4 we use the effective low-energy
theory to discuss the relationship between the strength of the $U(1)$
gauge coupling and the amount of ``tuning'' of the topcolor coupling
that is required.  In section 5 we show that, because the high-energy
theory includes a non-asymptotically free $U(1)$ gauge interaction, some
tuning is required.  By requiring that the Landau pole
\cite{Landau:1955} associated with the $U(1)$ gauge theory be
sufficiently above the topcolor gauge boson scale, we derive an upper
bound on the strength of the $U(1)$ gauge-coupling evaluated at the
topcolor scale.  The upper bound on the $U(1)$ coupling implies that the
topcolor coupling must be adjusted to equal the critical value for
chiral symmetry breaking to within a few per cent.

\section{A Puzzle in NJL}

The business end of a topcolor model \cite{Hill:1991at,Hill:1995hp}
consists of a set of quarks, including the $t$ and $b$, that transform
under topcolor, a stronger version of color $SU(3)$. There is also a
weaker $SU(3)$ gauge interaction that we will call ``protocolor''. In
the NJL approximation, the formation of composite Higgs bosons is driven
by a four-fermion operator obtained by integrating out the massive
colorons --- the gauge bosons of the strong topcolor interactions. The
mass, $M$, of the colorons comes from spontaneous breaking of topcolor
cross protocolor, which preserves ordinary color. We want to discuss this
spontaneous symmetry breaking in detail, so let us imagine for
simplicity that it comes from the VEV of a spinless field that
transforms like a $(3,\bar3)$ under topcolor cross protocolor. All the
essentials will be the same if the breaking is done in any simple
dynamical way. The coloron mass will then be given by
\begin{equation}
M^2=(g_{tc}^2+g_{pc}^2)\,v^2
\label{coloronmass}
\end{equation}
where $g_{tc}$ and $g_{pc}$ are the topcolor and protocolor couplings at the
scale $M$, $v$ is the VEV, and the coupling of ordinary color at the scale $M$
is then given by
\begin{equation}
g_c^2={g_{tc}^2\,g_{pc}^2\over g_{tc}^2+g_{pc}^2}
\end{equation}
which is approximately equal to $g_{pc}^2$ if $g_c^2\ll g_{tc}^2$. At low
energies, coloron exchange can be approximated by a
four-fermion coupling of the form
\begin{equation}
\half\,{
(g^2_{tc}\,j_{tc}^\mu-g^2_{pc}\,j_{pc}^\mu)
(g^2_{tc}\,{j_{tc}}_\mu-g^2_{pc}\,{j_{pc}}_\mu)\over
(g_{tc}^2+g_{pc}^2)^2\,v^2
}
\label{coloroninteraction}
\end{equation}
where $j_{tc}^\mu$ and $j_{pc}^\mu$ are the topcolor and protocolor currents.

There is something odd about the NJL approximation in a topcolor model.
Note that all dependence of (\ref{coloroninteraction}) on the topcolor
coupling goes away in the limit that
topcolor is much stronger than ordinary color --- which is a good
approximation at the topcolor scale. In this limit, 
the four-fermion interaction is nearly independent of the value of the
topcolor gauge coupling, because the coupling appears both in the numerator,
in the coupling to the top quark and other relevant fermions, and in the
denominator, in $M$. The way the NJL approximation deals with
this peculiarity is to treat $M$ as a fixed cut-off. Then what determines
whether
composite Higgs bosons are formed is the size of the four-fermion interaction
times the cut-off squared 
--- and this is proportional to the topcolor coupling $\alpha_{tc}$ evaluated
at the scale $M$. Thus in the NJL approximation, by adjusting the topcolor
coupling we can tune close to the critical ``coupling'' at which the composite
Higgs become massless. 

This peculiar behavior returns, however, when we now think about tilting
the vacuum with a $U(1)$ coupling. Typically \cite{Hill:1995hp}, such a
tilting interaction is also a spontaneously broken gauge interaction,
and in the NJL approximation is again described by a four-fermion
operator. The simplest possibility is to have the spinless field
or condensate that produces the topcolor breaking transform nontrivially
under the strong $U(1)$ as well. Again the details simplify further if we
imagine that this tilting $U(1)$ has a much larger coupling than any
other $U(1)$ in the theory. Then the low-energy four-fermion operator
due to $U(1)$ boson exchange looks like
\begin{equation}
\half\,{g_1^2\,j_1^\mu\,{j_1}_\mu\over m^2}
=\half\,{j_1^\mu\,{j_1}_\mu\over Q^2\,v^2}~,
\label{uoneinteraction}
\end{equation}
where $Q$ is the $U(1)$ charge of the field that breaks the symmetry,
$j_1^\mu$ is the $U(1)$ current, and $m$ is the $U(1)$ gauge boson mass,
satisfying
\begin{equation}
m^2\equiv g_1^2\,Q^2\,v^2\,.
\label{uonemass}
\end{equation} 

There are two peculiar things about (\ref{uoneinteraction}):
\begin{enumerate}
\item It does not depend on $g_1$.
\item It does depend on $Q$, and the interaction can be made very large by
taking $Q$ small (while leaving the couplings to the $t$- and
$b$-quarks fixed).
\end{enumerate}
Both of these features are physically unreasonable. Do we really
believe that the $U(1)$ interaction tilts the vacuum even in the
limit when its coupling $g_1$ is very tiny? Should we believe that for
moderate $g_1$ we can
still get very large tilting just by taking $Q$ and thus $m$ very small? We
think not, and indeed
these puzzling features do not persist in an effective field theory analysis
that goes beyond the NJL approximation.

\section{Matching in Large $N$}

We consider the effect of a spontaneously broken $U(1)$ on the effective
field theory describing the composite Higgs in a topcolor model near the
critical coupling. We will work in a large $N$ expansion
\cite{'tHooft:1974jz} for the topcolor interactions, but will not assume
that the NJL approximation is valid. We will assume instead that the low
energy theory is consistently defined in a mass independent
renormalization scheme such as dimensional regularization with modified
minimal subtraction ($DR\bar{MS}$).  We assume that the topcolor theory
can be matched at the scale $M$ (the mass of the colorons) onto
a theory of massless colored quarks coupled to light scalars ($\Phi$),
and study the effect of the massive $U(1)$ gauge boson. While the
matching at the scale $M$ is nonperturbative, below the scale $M$,
without the strongly coupled topcolor gauge bosons, the theory can be
analyzed perturbatively (until we get down to the QCD scale where
ordinary color gets strong).  After reviewing some simple consequences
of the $1/N$ expansion and effective field theory, we will discuss some
general properties of the matching that we can establish simply by
requiring that the physics be continuous as we vary the mass of the
$U(1)$ gauge boson.

\subsubsection*{An illustrative example --- the Noether current}

It is convenient to begin with a calculation that is trivial. Consider the
matching of a current,
\begin{equation}
j^\mu=\bar{\psi_L}\,\gamma^\mu\,Q_L\,\psi_L+
\bar{\psi_R}\,\gamma^\mu\,Q_R\,\psi_R
\label{jh}
\end{equation}
in the high energy theory onto the low energy theory. This is the conserved
Noether current associated with a $U(1)$ symmetry in the high energy theory.
It is therefore mapped onto the corresponding Noether current in the low
energy theory, which has the form:
\begin{equation}
\tilde j^\mu=\bar{\psi_L}\,\gamma^\mu\,Q_L\,\psi_L+
\bar{\psi_R}\,\gamma^\mu\,Q_R\,\psi_R
+i\tr\biggl[\Phi^\dagger\,\stackrel{\leftrightarrow}{\partial}^\mu\,
(Q_L\,\Phi-\Phi\,Q_R)\biggr]\,.
\label{jl}
\end{equation}
Note that the current $\tilde j^\mu$ in the low energy theory contains
contributions from
both the fermions and the composite Higgs, as it must, because both transform
under the $U(1)$ symmetry associated with the current.

\subsubsection*{$U(1)$ gauge boson exchange --- leading order in $N$}

Let us now consider the only slightly less trivial calculation of integrating
out a $U(1)$ gauge boson, which couples to the $L$ and $R$ topcolored fermions
with charges $Q_L$ and $Q_R$, respectively. If the $U(1)$ gauge boson's mass
is greater
than $M$, we integrate out the gauge boson before matching onto the low energy
theory to obtain the 4-fermion operator
\begin{equation}
{g_1^2\over2\,m^2}\,j^\mu\,j_\mu
\label{jjh}
\end{equation}
where $g_1$ is the coupling, $m$ is the $U(1)$ gauge boson mass and $j^m$ is
the current in (\ref{jh}). 
In leading order in $N$, the matching is trivial. Because the currents are
invariant under the topcolor gauge transformations, the leading contributions
factor \cite{Bardeen:1986vp,Chivukula:1986du}, and the result is\footnote{Note
that here, renormalization group running plays no role because both operators
are renormalized at the the same scale, $M$.} 
\begin{equation}
{g_1^2\over2\,m^2}\,\tilde j^\mu\,\tilde j_\mu
\label{jjl}
\end{equation}

This simple result contains an important lesson. In
general, the results of matching onto the low energy theory will involve both
the fermions and the scalars. We will come back to this momentarily.

If instead, the $U(1)$ gauge boson's mass is less than $M$, the gauge boson
survives into the low energy theory, where it now couples to the low energy
current $\tilde j^\mu$. Now when the gauge boson is integrated out, we again
obtain (\ref{jjl}). In this case, continuity of the physics across the
boundary $m=M$ is automatically satisfied and does not give us any more
information.

\subsubsection*{Matching a mass term}

Now consider the matching of a mass operator
\begin{equation}
\Sigma\equiv \bar{\psi_R}\,\psi_L\,.
\label{sigmah}
\end{equation}
In this case, we cannot evaluate the matching exactly, but we can use
symmetry arguments, $N$ and loop counting, and dimensional analysis to
write the dominant contribution in terms of a small number of parameters
of order one.\footnote{Another difference between this case and the case of
the
Noether current is that the mass operator has an anomalous dimension.
Such anomalous dimensions are very important for the structure of the
low energy theory at scales much smaller than $M$, because they give
rise to large logarithms from renormalization that allow us to calculate
some quantities reliably in spite of the nonperturbative nature of the
matching~\cite{Bardeen:1990ds}. Here, however, we will be discussing
matchings at or near the scale $M$, so there are no large logs.}
The matching produces an operator in the low energy theory of the form
\begin{equation}
\tilde\Sigma=A\,\bar{\psi_R}\,\psi_L
+B\,{\sqrt N\over4\pi}\,M^2\,\Phi
+C\,{4\pi\over\sqrt N}\,\Phi\,\Phi^\dagger\,\Phi
+\cdots
\,,
\label{sigmal}
\end{equation} 
where the $\cdots$ are higher dimension terms suppressed by powers of
$1/M$ in the low energy theory. The parameters $A$, $B$, and $C$ are of
order one (at the scale $M$), but their precise value depend on the
details of the strong topcolor physics.  The factors of ${\sqrt
  N\over4\pi}$ come from $N$ and loop counting. One can think about them
as follows. The coupling of $\Phi$ to the fermions must be of order
$4\pi/\sqrt N$, in order that the $\Phi$ kinetic energy term which comes
from planar diagrams like Fig.~\ref{fig-1} have the standard
normalization.

{\begin{figure}[htb]
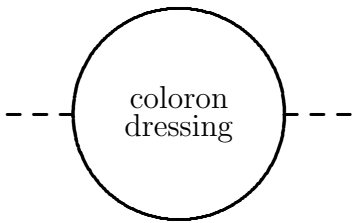

$$\beginpicture
\setcoordinatesystem units <\tdim,\tdim>
\stpltsmbl
\circulararc 360 degrees from 40 0 center at 0 0
\put {\stack{coloron,dressing}} at 0 0 
\setdashes
\plot 40 0 70 0 /
\plot -40 0 -70 0 /
\linethickness=0pt
\putrule from 0 50 to 0 -50 
\endpicture$$ 
\caption{\label{fig-1}\small Typical Feynman graph contributing to the $\Phi$
kinetic energy term. Such graphs, with arbitrary planar coloron dressing, are
proportional to ${N\over16\pi^2}$, and thus the $\Phi$ coupling should be of
order ${4\pi\over\sqrt{N}}$.}
\end{figure}}

With the result, (\ref{sigmal}), we can immediately write down the result for
matching of a 4-fermion operator of the form $\tr(\Sigma\,\Sigma^\dagger)$.
In leading order in the $1/N$ expansion, this just goes into
$\tr(\tilde\Sigma\,\tilde\Sigma^\dagger)$. This follows immediately from
factorization for large $N$. This result and (\ref{sigmal}) might seem
slightly strange to a reader steeped in the NJL approximation. In the NJL
approximation, (\ref{sigmal}) smacks of double counting, because the mass
operator $\Sigma$ plays much the same role as the scalar field $\Phi$. In
particular, one might worry that the $\tr(\Sigma\,\Sigma^\dagger)$ term in
$\tr(\tilde\Sigma\,\tilde\Sigma^\dagger)$ would produce additional
contributions to the $\Phi$ dependence from loop graphs like those shown in
Fig.~\ref{fig-2}. But in the low energy theory, these graphs vanish
identically. They are relevant in the NJL approximation because of the use of
a momentum space cut-off of order $M$. In some way, the use of the momentum
space cut-off is supposed to mock up the nonperturbative physics of the strong
coloron exchange near the critical coupling. But in our language, these
contributions (for zero momentum transfer through the loops) are simply
eliminated by $DR\bar{MS}$
{\begin{figure}[htb]
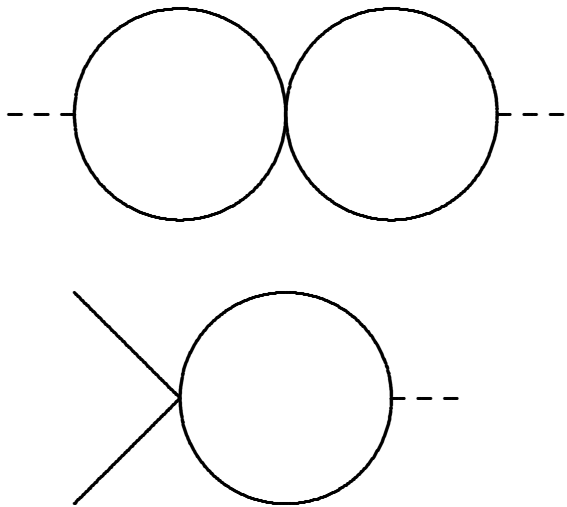

$$\beginpicture
\setcoordinatesystem units <\tdim,\tdim>
\stpltsmbl
\circulararc 360 degrees from 80 0 center at 40 0
\circulararc 360 degrees from -80 0 center at -40 0
\setdashes
\plot 80 0 110 0 /
\plot -80 0 -110 0 /
\linethickness=0pt
\putrule from 0 50 to 0 -50 
\endpicture$$ 
$$
\beginpicture
\setcoordinatesystem units <\tdim,\tdim>
\stpltsmbl
\circulararc 360 degrees from 40 0 center at 0 0
\plot -80 40 -40 0 -80 -40 /
\setdashes
\plot 40 0 70 0 /
\linethickness=0pt
\putrule from 0 50 to 0 -50 
\endpicture
$$
\caption{\label{fig-2}\small These graphs vanish in the low energy
theory at zero momentum transfer.}\end{figure}}

\subsubsection*{Dimensional Analysis}

As in any composite Higgs theory we can systematically estimate the size
of any term by dimensional analysis
\cite{Georgi:1993dw,Chivukula:1996sn}.  The intrinsic scale of the
compositeness interactions is the topcolor scale, $M$. Hence we expect
derivatives to be suppressed by a factor of $M$. If the amplitude for
the creation of a composite Higgs field is proportional to $f$ (the
analog of $f_\pi$ in chiral Lagrangian dimensional analysis in QCD
\cite{Weinberg:1979kz,Manohar:1984md}), each field $\Phi$ comes
suppressed by a factor of $f$. In a strongly interacting theory, we
expect there are no additional relevant parameters \cite{Georgi:1993dw}
in the low-energy theory. Requiring that the kinetic energy terms of the
fields $\Phi$ and $\psi$ have the canonical normalization then
determines that the overall size of each term in the Lagrangian is of
order $M^2 f^2$.

A priori, the dimensionless ratio $M/f$ is undetermined.  The effective
Lagrangian contains an infinite series of terms of arbitrary dimension.
As in the QCD chiral Lagrangian \cite{Weinberg:1979kz,Manohar:1984md},
loop diagrams involving lower dimension operators will require
counterterms of higher dimension.  Since the renormalization point
associated with the logarithms in the results of the loops is
undetermined to ${\cal O}(1)$, the {\it smallest} size of the
coefficients of these higher dimension operators is of order the typical
size of the coefficient of the chiral log for which the operator is a
counterterm. This consistency condition implies \cite{Chivukula:1992nw}
\beq
{M\over f} \sim {4\pi \over \sqrt{N}}~,
\label{nda}
\eeq
and the results given in the previous section immediately follow.  In
general, the $N$ counting implicit in (\ref{nda}) will give the
dependence of the {\it leading} contribution in large-$N$ and should
only be taken as a guide.  As we will see below, there are cases in which
the leading contribution vanishes and we will need to adjust the
$N$-dependence to correspond to the diagrams which would be evaluated in
matching the full and effective theories.

Using these rules to estimate the generic size of the composite Higgs
mass term in the effective theory, we find $m^2_\Phi \propto M^2$. This
is the hierarchy problem \cite{'tHooft:1980xb}.  In the absence of some
other symmetry not accounted for in these rules, some adjustment is
required to obtain $m^2_\Phi \ll M^2$.

\subsubsection*{$U(1)$ gauge boson exchange --- beyond leading order in $N$}

Now finally we are ready to discuss the effect of $U(1)$ gauge boson
exchange beyond the leading order in $N$. This is important, because it
is just such a nonleading effect that produces the ``tilting'' of the vacuum
alignment that is crucial to the phenomenology of topcolor
models~\cite{Hill:1991at,Hill:1995hp,Chivukula:1998}. We begin by discussing
the
situation for $m>M$ where we first integrate out the $U(1)$ gauge boson
to obtain the 4-fermion operator (\ref{jjh}). In particular, the
interesting effect involves both the left-handed and right-handed parts
of the currents:
\begin{equation}
{g_1^2\over m^2}\,\left[ \bar{\psi_L}\,\gamma^\mu\,Q_L\,\psi_L\right]\,
\left[ \bar{\psi_R}\,\gamma_\mu\,Q_R\,\psi_R\right]
\label{jjnl}
\end{equation}
This gives rise to symmetry-breaking matching contributions to the $\Phi$
mass and Yukawa coupling
from arbitrary planar coloron dressings of the diagrams in Figs.~\ref{fig-3}
and \ref{fig-4},
where the dotted line connects the left-handed and right-handed gauge
invariant parts of the 4-fermion operator in (\ref{jjnl}) (note that the
4-fermion contribution is already included in (\ref{jjl})). These give
following contributions to the $\Phi$ mass and Yukawa coupling
\begin{equation}
-D\,{g_1^2\,M^4\over 16\pi^2
m^2}\,\tr\biggl[Q_L\,\Phi\,Q_R\,\Phi^\dagger\biggr]
+E\,{g_1^2\,M^2\over 16\pi^2 m^2}\,{4\pi\over\sqrt
N}\,\bar{\psi_L}\,Q_L\,\Phi\,Q_R\,\psi_R+{\rm h.c.}+\cdots
\label{jjnoh}
\end{equation}
where $D$ and $E$ are of order 1.\footnote{In the NJL approximation, $D$
is positive.} Naive application of dimensional analysis would yield an
additional factor of $N$.  However, these contributions (see
Fig.~\ref{fig-3} and \ref{fig-4}) are formally nonleading in $N$ because
the fermion lines connect the two gauge invariant parts of the operator
(\ref{jjnl}).  This can also be seen directly from the diagrams: the
factors of $1/\sqrt{N}$ in the $\Phi$ coupling cancel against the $N$ from the
single fermion loop. Note that ``planar'' in this case includes
graphs in which the coloron lines go through the dotted line in
Fig.~\ref{fig-3} and \ref{fig-4}. Thus this contribution does not
factorize, and we cannot relate it to (\ref{sigmal}). Note also that the
$U(1)$ gauge coupling goes like $1/\sqrt N$ for large
$N$~\cite{Chivukula:1998} so the contribution of (\ref{jjnoh}) to the
$\Phi$ mass goes to zero as $N\rightarrow\infty$.

{\begin{figure}[htb]
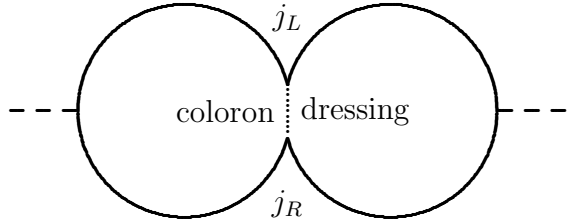

$$\beginpicture
\setcoordinatesystem units <\tdim,\tdim>
\stpltsmbl
\circulararc -331 degrees from 0 10 center at 39 0
\circulararc 331 degrees from 0 10 center at -39 0
\put {$j_L$} [b] at 0 30
\put {$j_R$} [t] at 0 -30
\put {coloron} [r] at -5 0
\put {dressing} [l] at 5 0 
\setdashes
\plot 80 0 110 0 /
\plot -80 0 -110 0 /
\setdots <2pt>
\plot 0 10 0 -10 /
\linethickness=0pt
\putrule from 0 50 to 0 -50 
\endpicture$$ 
\caption{\label{fig-3}\small 
Symmetry breaking matching contribution to the $\Phi$ mass. The dotted line
connects
two gauge invariant parts of the 4-fermion operator. The matching contribution
incorporates the effect of the coloron dressing which is not present in the
low energy theory. }\end{figure}}
{\begin{figure}[htb]
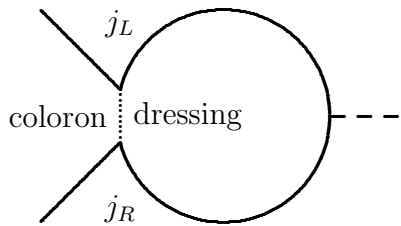

$$\beginpicture
\setcoordinatesystem units <\tdim,\tdim>
\stpltsmbl
\circulararc -331 degrees from 0 10 center at 39 0
\plot 0 10 -30 40 /
\plot 0 -10 -30 -40 /
\put {$j_L$} [b] at 0 30
\put {$j_R$} [t] at 0 -30
\put {coloron} [r] at -5 0
\put {dressing} [l] at 5 0 
\setdashes
\plot 80 0 110 0 /
\setdots <2pt>
\plot 0 10 0 -10 /
\linethickness=0pt
\putrule from 0 50 to 0 -50 
\putrule from -30 0 to 110 0 
\endpicture$$ 
\caption{\label{fig-4}\small 
Symmetry breaking matching contribution to the $\Phi$ Yukawa coupling. The
dotted line
connects two gauge invariant parts of the 4-fermion operator. The matching
contribution incorporates the effect of the coloron dressing which is not
present in the low energy theory.}\end{figure}}

Note that the diagrams in Figs.~\ref{fig-3} and \ref{fig-4} are really the
same graphs as those in Fig.~\ref{fig-2}, except that they contain coloron
dressing. It is the coloron effects that produce the matching
contribution, and the scale $M$ that is required to give a nonzero result.

Now what happens if $m<M$? Now there are a number of contributions that arise
from the diagrams in Figs.~\ref{fig-5} and \ref{fig-6}, both with and
without coloron dressing. There are effects from integrating out the colorons
at the scale $M$; these
contributions to the $\Phi$ mass and Yukawa coupling have the form
\begin{equation}
-D'\,{g_1^2\over 16\pi^2}\,M^2\,\tr\biggl[Q_L\,\Phi\,Q_R\,\Phi^\dagger\biggr]
+E'\,{g_1^2\over 16\pi^2}\,{4\pi\over\sqrt
N}\,\bar{\psi_L}\,Q_L\,\Phi\,Q_R\,\psi_R+{\rm h.c.}+\cdots\,.
\label{jjnol}
\end{equation}
There also are effects from
from integrating out the massive $U(1)$ gauge boson at the scale $m$ in the
low
energy theory; these
contributions to the $\Phi$ mass and Yukawa coupling have the form
\begin{equation}
-D''\,{g_1^2\over 16\pi^2}\,m^2\,\tr\biggl[Q_L\,\Phi\,Q_R\,\Phi^\dagger\biggr]
+E''\,{g_1^2\over 16\pi^2}\,{4\pi\over\sqrt
N}\,\bar{\psi_L}\,Q_L\,\Phi\,Q_R\,\psi_R+{\rm h.c.}+\cdots\,.
\label{jjnolm}
\end{equation}
where the parameters $D'$, $E'$, $D''$ and $E''$ are of order 1. In addition,
there are
calculable effects from the renormalization group running in the low energy
theory that causes these coupling to evolve with the energy scale. The
interesting difference between (\ref{jjnoh}) in the high energy theory and
(\ref{jjnol}) and (\ref{jjnolm}) in the low energy theory
is the $1/m^2$ dependence in (\ref{jjnoh}). There is no way that either the
coloron effect at the scale $M$, or the running and matching below $M$ can
produce the $1/m^2$ dependence. At $M$, the typical momentum in the matching
calculation is of order $M$, so the $U(1)$ gauge boson propagator does not
produce a $1/m^2$. Below $M$, we can never get an $m^2$ in the denominator of
the mass term or Yukawa coupling (or the $\Phi^4$ coupling, which we have not
written), because it would have to be compensated by factors of $M^2$ in the
numerator, and
there is no way that such factors can arise in the low energy theory. They can
originate in the matching at the scale $M$ from the nonperturbative physics of
the coloron exchange, but new factors of $M$ cannot occur in the low energy
theory.

{\begin{figure}[htb]
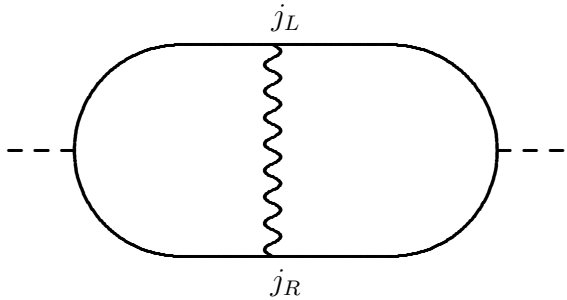

$$\beginpicture
\setcoordinatesystem units <\tdim,\tdim>
\stpltsmbl
\circulararc -180 degrees from 40 40 center at 40 0
\circulararc 180 degrees from -40 40 center at -40 0
\plot 40 40 -40 40 /
\plot 40 -40 -40 -40 /
\put {$j_L$} [b] at 0 45
\put {$j_R$} [t] at 0 -45
\photondr -5 40 *7 /
\setdashes
\plot 80 0 110 0 /
\plot -80 0 -110 0 /
\linethickness=0pt
\putrule from 0 50 to 0 -50 
\endpicture$$ 
\caption{\label{fig-5}\small 
Symmetry breaking contribution to the $\Phi$ mass in the low energy theory.
}\end{figure}}

{\begin{figure}[htb]
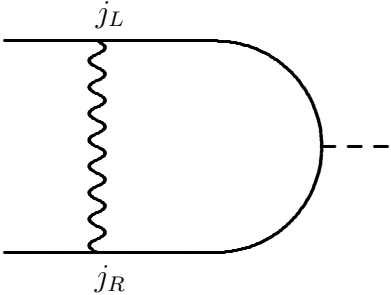

$$\beginpicture
\setcoordinatesystem units <\tdim,\tdim>
\stpltsmbl
\circulararc -180 degrees from 40 40 center at 40 0
\plot 40 40 -40 40 /
\plot 40 -40 -40 -40 /
\put {$j_L$} [b] at 0 45
\put {$j_R$} [t] at 0 -45
\photondr -5 40 *7 /
\setdashes
\plot 80 0 110 0 /
\linethickness=0pt
\putrule from 0 50 to 0 -50 
\putrule from -40 0 to 110 0 
\endpicture$$ 
\caption{\label{fig-6}\small 
Symmetry breaking contribution to the $\Phi$ Yukawa coupling in the low energy
theory. }\end{figure}}

Now we can impose the constraint that the physics should be continuous at the
boundary $m=M$. Clearly, this implies 
\begin{equation}
D=D'+D''\,,\quad\quad E=E'+E''\,.
\label{boundary}
\end{equation}
This determines the Yukawa coupling below the scale $m$ because it is only the
sum $E'+E''$ that contributes.

In fact, though we cannot prove it using these methods, we think it likely
that the most important symmetry breaking contribution to the $\Phi$ mass
comes from integrating 
out the $U(1)$ gauge boson, in which case we would conclude
\begin{equation}
D''\stackrel{?}{\gg}D'\,.
\label{maybe}
\end{equation}
This would have the
interesting consequence that the $U(1)$ interaction is most efficient at
tilting the vacuum when $m=M$. Then the contribution to the symmetry breaking
mass of the $\Phi$ would be
\begin{equation}
\Delta m_\Phi^2\propto \left\{\begin{array}{ll}
m^2 &\mbox{for $m<M$, and}\\
\displaystyle {M^4\over m^2} &\mbox{for $m>M$.}
\end{array}\right.
\label{deltam}
\end{equation}
But whether we can erase the question mark in (\ref{maybe}) or not, 
it is clear that the tilting effect of a $U(1)$ gauge boson as a function of
$m$ cannot be much larger that the effect at $m=M$.

We have now resolved the NJL puzzles posed in the previous section.  When
$g_1^2$ is small, all the contributions to tilting are proportional to
$g_1^2$, and we cannot get large tilting by taking $Q$ and thus $m$ very
small.

\section{Tuning and Tilting}

Now we apply the results of the previous section to discuss the
issue of fine tuning in topcolor models. The fine tuning we have in mind is
the tuning required to make the composite Higgs bosons much lighter than the
scale $M$. First, let us turn off the $U(1)$ interaction. Then the scalar
fields described by the field $\Phi$ have a common mass, $\mu$. For
sufficiently small $\mu$, we expect $\mu$ to be determined by a relation of
the form
\begin{equation}
\mu^2=F\, \left({\alpha_c-\alpha_{tc}(M)
\over\alpha_c
}\right)\,M^2\,,
\label{critical}
\end{equation}
where $F$ is a positive constant of order 1 and $\alpha_c$ is the critical
value
of the topcolor coupling for chiral symmetry breaking. In the NJL
approximation, we can compute $F$ and $\alpha_c$. In the more general
effective field theory description, we cannot calculate them reliably,
but the form of the relation should be true more generally so long as
the chiral symmetry breaking transition is second order.  We will say
that this is a fine tuning of order \cite{Chivukula:1995dc}
\begin{equation}
T={\alpha_c-\alpha_{tc}(M)
\over\alpha_c
}\,.
\label{finetuning}
\end{equation}

We now consider the $U(1)$ interaction.
For simplicity, we will consider the simple situation in which only the $t$
and $b$ carry topcolor. Then the field $\Phi$ consists of four light complex
scalar fields
which have the quantum numbers of two independent 2-component ``Higgs''
fields $(\phi_t\ \&\ \phi_b)$. The issue then is whether we can tilt the
vacuum so that the $\phi_t$ mass is negative and the $\phi_b$ mass is
positive, so that a top condensate will produce a large $t$ mass,
but no $b$ condensate will form.

Consider a model with $U(1)$-charge $y_L$ for the
left-handed $t-b$ doublet, and charges $y_{t_R}$ and $y_{b_R}$ for the
right-handed $t$ and $b$, respectively. 
Using the results of the previous section, we conclude that the contribution
of the $U(1)$ to the $\Phi$ mass cannot be larger than
\begin{equation}
-D\,{g_1^2\over 16\pi^2}\,M^2\,\tr\biggl[Q_L\,\Phi\,Q_R\,\Phi^\dagger\biggr]
\label{massdif1}
\end{equation}
for $D$ of order 1. In terms of the $U(1)$ charges, this
can now be written as
\begin{equation}
-D\,{g_1^2\over 16\pi^2}\,M^2\,\tr\left[y_L\,\pmatrix{
\phi_t&\phi_b\cr
}\,\pmatrix{
y_{t_R}&0\cr
0&y_{b_R}\cr
}\,\pmatrix{
\phi_t^\dagger\cr
\phi_b^\dagger\cr
}\right]
\label{massdif2}
\end{equation}
Combining (\ref{massdif2}) with the common mass $\mu$ gives mass terms for
$\phi_t$ and $\phi_b$ of order
\begin{equation}
m^2_{\phi_b}\approx\mu^2-D\,{g_1^2 \over 16\pi^2}\,M^2\,y_L\,y_{b_R}\,
,\quad\quad
m^2_{\phi_t}\approx\mu^2-D\,{g_1^2 \over
16\pi^2}\,M^2\,y_L\,y_{t_R}\,.
\label{massdif3}
\end{equation}

To produce a $t$ condensate, but no $b$ condensate, we must have a negative
mass squared for $\phi_t$ and a positive mass squared for $\phi_b$, thus we
want
\begin{equation}
\mu^2-D\,{g_1^2\over 16\pi^2}\,M^2\,y_L\,y_{b_R}>0\,,
\quad\quad
\mu^2-D\,{g_1^2 \over 16\pi^2}\,M^2\,y_L\,y_{t_R}<0\,,
\label{massdif4}
\end{equation}
which using (\ref{critical}) and (\ref{finetuning}) can be written as
\begin{equation}
{D\over F}\,{g_1^2\over 16\pi^2}\,y_L\,y_{b_R}<  T <
{D\over F}\,{g_1^2\over 16\pi^2}\,y_L\,y_{t_R}\,.
\label{bound1}
\end{equation}
This relation says that the larger $T$ is --- that is the less severe
the fine tuning to the critical coupling --- the larger $g_1^2$ (times
the charges) must be. $D/F$ is unknown, but is expected to be ${\cal
  O}(1)$. In the appendix, we show that $D/F = 4$ in the NJL
approximation.

\section{Some Tuning Required...}

We will now show that $T$ cannot be too large.  Because the high-energy
theory includes a non-asymptotically free $U(1)$ coupling, it too must
be only an effective theory below some cutoff energy $\Lambda$.  The
scale $\Lambda$ must be lower than the energy scale where the $U(1)$
gauge theory becomes strongly coupled -- {\it i.e.} lower than the
potential Landau pole \cite{Landau:1955}.  Consider the one-loop
$\beta$-function
\beq
\beta = \mu {dg_1(\mu)\over d\mu} =  {b\, g^3_1 \over 24 \pi^2} + \ldots
\eeq
where $b= \sum_f y^2_f$ and the sum runs over all left-
and right-handed $U(1)$ charges. Using the first term as
an estimate, we find an upper bound on the strength
of the $U(1)$ coupling at scale $M$
\beq
g_1(M) \laem {12 \pi^2 \over b\, \log{\Lambda\over M}} ~,
\eeq
From eqn. (\ref{bound1}) we then obtain
\beq
|T|  < {3\,|y_L|\, |y_R|^{\rm max} \over 4\,b\, \log\left({\Lambda\over
M}\right)}
\, {D \over F}~,
\label{tbound}
\eeq
where $|y_R|^{\rm max} = {\rm max}(|y_{t_R}|,|y_{b_R}|)$.  Note that
this relation makes physical sense: only the {\it ratios} of
$U(1)$-charges appear.

Since the top- and bottom-quarks {\it must} couple to the $U(1)$, we
know
\beq
b \gaem N\,(2|y_L|^2 + (|y_R|^{\rm max})^2).
\eeq
The ratio of $U(1)$ charges in (\ref{tbound}) has
a maximum value:
\beq
{|y_L| |y_R|^{\rm max} \over 2|y_L|^2 + (|y_R|^{\rm max})^2}
\le {1\over 2\sqrt{2}}~.
\eeq
This yields the model-independent bound
\beq
T < {\cal O}(4\%)
\eeq
for $\Lambda/M \ge 10$. That is, the topcolor coupling renormalized at
scale $M$ must be tuned equal the critical value for chiral symmetry
breaking to $\sim 4\%$! Furthermore, this implies that the $U(1)$
coupling cannot be particularly strong: it can provide a shift of only
${\cal O}(4\%)$ of $M^2$ to the composite Higgs boson mass-squared.

Stronger bounds can be obtained in specific models.  In the ``minimal''
topcolor assisted technicolor model \cite{Hill:1995hp}, the strong
$U(1)$ couples only to the third generation of quarks and leptons with
charges proportional to hypercharge. Here $b = 40/3$ and one finds $T
<{\cal O}(1\%)$. This can severely restrict the relevant parameter
space in topcolor models (for example, see \cite{Popovic:1998}).

In a more realistic model, which tries to accommodate intergenerational
mixing, prevent large amounts of isospin violation, and avoid the
presence of light pseudo-Goldstone bosons \cite{Lane:1996ua,Lane:1997},
the number of fermions coupling to the $U(1)$ is likely to be much
larger. As noted by \cite{Lane:1996ua}, the constraints will likely
be even stronger.  For example from eqn.~(18) of ref. \cite{Lane:1997} (with
the additional assumption that $z_1 = 1.0$) we find $b \approx 117$ and
$y_L \approx y^{\rm max}_R \approx 1$. This yields $T < {\cal
  O}(0.2\%)$.

\section{Conclusions}

In this paper we have investigated the chiral-symmetry breaking dynamics
in models of topcolor and topcolor-assisted technicolor which rely on a
relatively strong $U(1)$ gauge interaction to ``tilt'' the vacuum.  We
identified some peculiarities of the Nambu-Jona-Lasinio (NJL)
approximation often used to analyze the topcolor dynamics. We then
resolved these puzzles by constructing the low-energy effective field
theory appropriate to a mass-independent renormalization scheme and by
constructing the power-counting rules appropriate to such an effective
theory.  Requiring that the Landau pole associated with the $U(1)$ gauge
theory be sufficiently far above the topcolor gauge boson scale, we derived
an upper bound on the strength of the $U(1)$ gauge-coupling evaluated at
the topcolor scale. The upper bound on the $U(1)$ coupling implies that
this interaction can only shift the composite Higgs boson mass-squared by a
few
per cent and, therefore, that the topcolor coupling must be adjusted to
equal the critical value for chiral symmetry breaking to within a few
per cent.

In much of this work the assumption that $N$ is large does not play a
crucial role. It restricts the form of various terms, but the basic
structure of the effective field theory arguments remains the same,
assuming that the transition remains second
order~\cite{Chivukula:1993pm,Bardeen:1994pj}. The one place where large
$N$ is important is in guaranteeing that it is the strong topcolor
interactions that are driving the formation of the composite Higgs
state, $\Phi$, while the $U(1)$ interaction acts as a $1/N$
perturbation~\cite{Chivukula:1998}. Because of asymptotic freedom, there
is a plausible physical picture of the critical transition driven by
topcolor. It results from the fact that the strong topcolor interactions
get strong at a scale $\Lambda_{tc}\approx M$.

If $N$ cannot be treated as large, then perhaps this can be turned on
its head, but we admit that we do not understand the very different
scenario in which the $U(1)$ interaction is nearly critical and the
topcolor interaction is weak.  It is much less obvious what the
effective field theory of a nearly critical U(1) is like --- if such a
thing exists at all.  We have nothing to say about this, except that
there is no chance of it's making sense unless $N$ is small.

\bigskip


\centerline{\bf Acknowledgments}

We thank Elizabeth Simmons for comments on the manuscript.  {\em This
  work was supported in part by the Department of Energy under grant
  DE-FG02-91ER40676 and the National Science Foundation under grant
  PHY-9218167.}


\appendix

\subsection*{Appendix: Tuning in the NJL Approximation}

Consider a model with $U(1)$-charge $y_L$ for the
left-handed $t-b$ doublet, and charges $y_{t_R}$ \& $y_{b_R}$ for the
right-handed $t$ and $b$ respectively. In the NJL approximation,
assuming the masses of the topcolor and hypercharge gauge bosons are
comparable, the interactions of the third generation of quarks are
approximated by the four-fermion operators
\beq
{\cal L}_{4f} = -{{g^2_{tc}(M)}\over{ 2\, M^2}}\left[\overline{\psi}\gamma_\mu
{{\lambda^a}\over{2}}
\psi \right]^2
-{{g^2_1(M)}\over{2\,M^2}}\left[y_L \overline{\psi_L}\gamma_\mu  \psi_L
+y_{t_R} \overline{t_R}\gamma_\mu  t_R
+y_{b_R} \overline{b_R}\gamma_\mu  b_R
\right]^2~.
\label{L4t}
\eeq
Here $\psi$ represents the top-bottom doublet, and the $g^2_{tc}(M)$ and
$g^2_1(M)$ are respectively the top-color and $U(1)$ gauge-couplings
squared evaluated at scale $M$.  The usual NJL gap-equation condition
\cite{Buchalla:1996dp} for top-condensation but not bottom-condensation
is
\beq
g^2_{tc}(M)+{2\over N} y_L y_{b_R} g^2_1(M)
\laem g^2_c={8\pi^2\over N} \laem
g^2_{tc}(M)+{2\over N} y_L y_{t_R} g^2_1(M)~.
\eeq
From this we see that
\beq
{g^2_1(M) \over 4\pi^2} y_L y_{b_R} \laem T \laem {g^2_1(M) \over 4\pi^2} y_L
y_{t_R}~.
\eeq
This corresponds to eqn. (\ref{bound1}) with $D/F = 4$.

\bibliography{tilting}
\bibliographystyle{h-elsevier}

\end{document}